\documentclass[9pt,a4paper]{article}

\usepackage{tocloft}
\usepackage[british]{babel}

\usepackage[margin=0.6in]{geometry}

\usepackage[style=numeric]{biblatex}
\usepackage{amsmath}
\usepackage{graphicx}
\usepackage[colorlinks=true, allcolors=blue]{hyperref}
\usepackage{hyperref}
\usepackage[title]{appendix}
\usepackage{mathrsfs}
\usepackage{amsfonts}
\usepackage{booktabs} 
\usepackage{caption}  
\usepackage{threeparttable} 
\usepackage{algorithm}
\usepackage{algorithmicx}
\usepackage{algpseudocode}
\usepackage{listings}
\usepackage{enumitem}
\usepackage{chngcntr}
\usepackage{booktabs}
\usepackage{lipsum}
\usepackage{subcaption}
\usepackage{authblk}
\usepackage[T1]{fontenc}    
\usepackage{csquotes}       
\usepackage{diagbox}
\usepackage{cancel}

\usepackage{setspace}
\usepackage{titlesec}
\titleformat{\section} 
  {\normalfont\Large\bfseries}{\thesection.}{1em}{}
  
\usepackage{float}   
\usepackage{caption} 



\title{Susceptibility of entanglement entropy : \\
a universal indicator of quantum criticality}

\author{Pritam Sarkar}
\affil{Indian Association for the Cultivation of Science, Kolkata}
\affil[*]{email: \texttt{[prikarsartam@gmail.com]}}

\date{Nov 2024}  

\begin{document}
\maketitle

\begin{abstract}
A measure of how sensitive the entanglement entropy is in a quantum system, has been proposed and its information geometric origin is discussed.  It has been demonstrated for two exactly
solvable spin systems, that thermodynamic criticality is directly \textit{indicated} by finite size scaling of the global maxima and  turning points of the susceptibility of entanglement entropy through numerical analysis - obtaining power laws. Analytically we have proved those power laws for $| \ \lambda_c(N)-\lambda_c^{\infty}|$ as $N\to \infty$ in the cases of 1D transverse field ising model (TFIM) ($\lambda=h$) and XY chain ($\lambda=\gamma$). The integer power law appearing for XY model has been verified using perturbation theory in $\mathcal{O}(\frac{1}{N})$ and the fractional power law appearing in the case of TFIM, is verified by an exact approach involving Chebyshev polynomials, hypergeometric functions and complete elliptic integrals. Furthermore a set of potential applications of this quantity under quantum dynamics and also for non-integrable systems, are briefly discussed. The simplicity of this setup for understanding quantum criticality is emphasized as it takes in only the reduced density matrix of appropriate rank.
\end{abstract}

\tableofcontents

\section{Introduction  }

Highest sensitivity of global properties near some points in the parameter space is the general feature of a critical phenomenon, classical or quantum. Various theoretical methods of determining the nature of criticality in quantum many-body systems have been previously approached such as order parameter \cite{Heyl2017SpeedLimits}, renormalization group theory \cite{Tsai2001DMRG}, fidelity susceptibility \cite{damski, polkovnikov}, concurrence and entanglement measures \cite{Nielsen2005Entanglement,ZhukovEvolutionOfSingleSiteEntanglement, LarssonSingleSiteEntanglement} to name a few,  each with their own model-specific pros and cons.

Finite size scaling of fidelity susceptibility or more generally the geometric tensor \cite{damski, Kolodrubetz_2013, Zanardi2007Geometry, Zanardi2006Fidelity} has been previously shown to reflect the universality of quantum critical phenomenon in spin chains and free-fermion systems. Ground state fidelity naturally defines a riemannian metric in the parameter space \cite{Kolodrubetz_2013, Zanardi2007Geometry,Zanardi2006Fidelity}. Its turning points and value at the critical point is shown to be power-laws in system size for transverse field Ising model \cite{damski} and free-fermions \cite{Zanardi2006Fidelity}. This requires integrability of the theory to analytically obtain the ground state or approximating the susceptibility of the ground state fidelity using adiabatic gauge potential \cite{polkovnikov, Kolodrubetz_2013}.

Here we propose a novel model-agnostic tool to determine the susceptibility of entanglement entropy for a quantum many-body system inspired from geometric interpretation of fidelity susceptibility and ground state manifold \cite{polkovnikov, Kolodrubetz_2013, Zanardi2006Fidelity}. We demonstrated how its finite-size scaling directly indicates the thermodynamic criticality through the convergence of its turning-points and divergence or saturation of its global maximum. 'Universal' bears a two-fold meaning here, firstly as a method that can be used across different models and as a quantity containing the scaling-information of a certain universality class. The robustness and generalisability of this susceptibility of entanglement entropy lies in its origin in information geometry \cite{nielsen, EunJinKimGeodNonEqComplex,Kim2011InfoGeometry} as it is formally diagonal elements of the quantum analog of Fisher-Rao information metric. The free fermionic exact solution for a XY chain with transverse magnetic field and periodic boundary is used \cite{Lieb1961SolubleModels,Sachdev2011QPT} to establish the results numerically and verified analytically for two different models, XY spin chain and transverse field Ising chain (TFIM) with a closed-form asymptotic expression of ground state energy density and transverse magnetization of TFIM with interesting intervention of several special functions. This work is organised as follows, starting with its derivation using an idea from information geometry the exact solution of the model is used to formulate the susceptibility of entanglement entropy of the model, followed by numerical results and analytical explanations for two spin chains with different local interactions, ending with potential ramifications in broader applications.









\section{Information geometric origin }

A concrete notion of \textit{how different two states are} can be defined by a distance between the states at their respective parameter values. This can be derived from the relative entanglement entropy between density matrices of at infinitesimally separated points in the parameter space and that satisfies all conditions of a riemannian metric. It is closely related to \textit{Fisher-Rao information metric} as studied in non-equilibrium classical physics \cite{Kim2011InfoGeometry, nielsen, EunJinKimGeodNonEqComplex}.
\begin{equation}
      S(\hat{\rho} || \hat{\sigma}) = \text{Tr}[ \hat{\rho} ( ln \hat{\rho} - ln  \hat{\sigma} )] \ \ \implies  S(\hat{\rho}_{\lambda} || \hat{\rho}_{\lambda+\delta \lambda})  = \Sigma_{ij}(\vec{\lambda}) \ d \lambda^i \ d \lambda^j \ + \ \mathcal{O}(\delta \lambda^3), \ \ \ \text{with} 
\end{equation}
\begin{equation}
    \Sigma_{ij}  = \frac{1}{2} \ \text{Tr}[ \hat{\rho} \ \partial_i(ln \ \hat{\rho}) \ \partial_j(ln \ \hat{\rho})] \  \label{GenSusceptEntEnt}
\end{equation}
For models with one parameter, we just need to consider the other parameters as constants and take the corresponding diagonal $\Sigma_{ii}$ as \textit{susceptibility of entanglement entropy w.r.t. the parameter $\lambda_i$}. This is similar to the relation between fidelity susceptibility and geometric tensor \cite{polkovnikov}. But without directly relying on integrability or any approximation scheme density matrices can be exactly computed or analytically computed through various methods \cite{Barouch1970StatMechI, Barouch1971StatMechII} . It allows this indicator to be model-agnostic.  

To the best of our knowledge no study has been made before about finite size scaling of quantum critical phenomenon, from the perspective of how sensitive entanglement entropy is, using information theory.

\section{Finite XY chain in transverse magnetic field }
The exactly solvable model we will discuss is given by the hamiltonian \cite{Barouch1970StatMechI, polkovnikov, Lieb1961SolubleModels, Sachdev2011QPT}: 
\begin{equation}
    \mathcal{H} = - \sum_{j=1}^{L}  \left( \frac{1 + \gamma}{2} \right)  \ \sigma_j^x \sigma_{j+1}^x + \left( \frac{1 - \gamma}{2} \right)  \ \sigma_j^y \sigma_{j+1}^y - h \ \sigma_j^z  \ 
    \label{eq:PrimaryHam__0}
\end{equation}
which can be mapped into a free-fermionic theory after Jordan-Wigner-Fourier-Bogolioubov transformation and considering periodic boundary condition \cite{Jordan1928Wigner, Sachdev2011QPT, polkovnikov}. 
\begin{equation}
\mathcal{H} =     - \sum_k \Psi_k^{\dagger} \begin{pmatrix}  (h-cosk) & -\gamma sink \ \\  
-\gamma sink \  & -(h-cosk)   
\end{pmatrix} \Psi_k   = \sum_k \epsilon_k \ \gamma^\dagger_{k} \gamma_{k} + \text{Const} \ \ \  \text{with: } \  
\label{XYChainWithTransverseField}
\end{equation}
\begin{equation*}
\Psi_k =\begin{pmatrix}  c_k \\  c^\dagger_{-k}  \end{pmatrix}= \begin{pmatrix} \ cos\frac{\theta_k}{2} & i \ sin\frac{\theta_k}{2} \\  i \ sin\frac{\theta_k}{2} & cos\frac{\theta_k}{2} \end{pmatrix} \begin{pmatrix} \ \gamma_k \\  \gamma^\dagger_{-k} \end{pmatrix}, \ \ \theta_k = \tan^{-1}(\frac{\gamma sink}{h - cosk}),  \ \ \ \& \ \ \ \epsilon_k = \pm \sqrt{(h-cosk)^2 + \gamma^2 sin^2k}, 
\end{equation*}
In this fermionic language the ground state is given as a spinless BCS ground state \cite{polkovnikov, Sachdev2011QPT}: 
\begin{equation}
    | \ 0 \ \rangle = \bigotimes_{k} ( \ cos\frac{\theta_k}{2} - sin\frac{\theta_k}{2} \ c^{\dagger}_{k} c^{\dagger}_{-k}\ ) | 0_k \rangle, \ \ \text{such that: } \ \langle 0 | 0 \rangle = 1, \ \ \& \ \ \gamma_k | 0 \rangle = 0 \label{BCSgroundState}
\end{equation}
\section{Susceptibility of entanglement entropy}
 Now we require a density matrix that is easy to compute but gives sufficient information about the critical features. Since the order parameter in this theory is the expectation value of a single-body operator, the \textit{1-body reduced density matrix} $\hat{\rho}^{(1)}$ is enough for our purpose. Using operator product expansion we know 
$$\hat{\rho}^{(1)} = \frac{1}{2} (\ \mathbb{I}_{2 \times 2} + \sum_{\alpha \in \{ x, y, z\}} \text{Tr}[\hat{\rho} \ \hat{\sigma}^{\alpha}] \ \hat{\sigma}^{\alpha} \ ), \  \ \ \text{with the full density matrix: } \ \hat{\rho} = \frac{e^{ - \beta \hat{\mathcal{H}}}}{\text{Tr}[e^{ - \beta \hat{\mathcal{H}}}]}$$

In our context, $\beta \to \infty$ as we are looking at the pure quantum state at $T \to 0$ with no thermal fluctuations. Furthermore, we will focus the analysis on finite systems to determine how its turning points, maximum susceptibility and FWHM (Full-Width at Half Maxima) varies as $N$, in two spin-chains with periodic boundaries, i.e. transverse field ising model and XY model with PBC.
\\
Note that the hamiltonian possesses the following symmetry : $\{ \hat{\sigma^{x}} \to -\hat{\sigma^{x}}, \ \hat{\sigma^{y}} \to - \hat{\sigma^{y}}, \ \hat{\sigma^{z}} \to \hat{\sigma^{z}} \}$, which is also maintained by the ground state of every finite system, except only the one with infinitely many spins. This makes expectation of $\hat{\sigma^{x}}$ and $\hat{\sigma^{y}}$ identically vanish in any finite system. 
\begin{equation}
\implies \hat{\rho}^{(1)} = \frac{1}{2} (\ \mathbb{I}_{2 \times 2} + m_z \ \hat{\sigma}^{z} \ ) \ \ , \ \ \ \text{where: } \ m_{z}(h,\gamma) =  \langle \ \sigma^{z} \ \rangle = \frac{1}{N} \sum_i \langle 0 | \sigma^z_i |0 \rangle = \frac{1}{N} \sum_k \frac{h-cosk}{\sqrt{(h-cosk)^2+\gamma^2 sin^2k} }\label{singleSiteReduceddensityMatrix}
\end{equation}
With the ground state $| 0 \rangle$ is given by \eqref{BCSgroundState}. More explicitly, Having $L$ equidistant angles (momentum modes) within $[-\pi, \pi]$  requires \cite{Barouch1970StatMechI} 
\begin{equation}
    m_z (h, \gamma, N)= \frac{1}{L}\sum_{l=-(\frac{N-1}{2})}^{l=(\frac{N-1}{2})} \frac{h-cos(\phi_l)}{\sqrt{(h-cos(\phi_l))^2+\gamma^2 sin^2(\phi_l)} }, \ \ \text{with: } \ \phi_l = \frac{2 \pi l}{N} \label{m_z__N}
\end{equation}
\begin{figure}[ht]
  \centering
  \includegraphics[width=\textwidth]{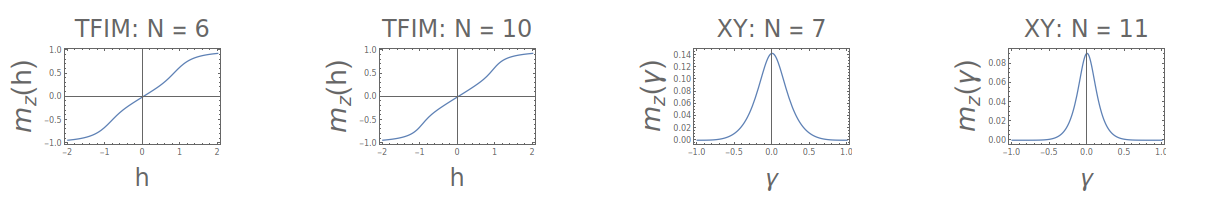}
  \caption{Examples of transverse magnetization for different system sizes}
  \label{fig:image132654}
\end{figure}
This completes the analytic expression for $\hat{\rho}^{(1)}$ that we can compute for any finite system size as defined in Eq. \eqref{singleSiteReduceddensityMatrix}. We are now ready to compute the susceptibility of entanglement-entropy of two systems at two different limits of this hamiltonian. 
  \begin{equation}
    \text{\textbf{ For XY model: }} \Sigma_{N}^{\text{XY}}(\gamma) = \Sigma_{\gamma \gamma}^{(h=0)}(N), \ \  \&  \label{suscept_Ent_Ent}
    \ \ \ \text{\textbf{ For transverse field Ising model : }} \ \Sigma_{N}^{\text{TFIM}}(h) = \Sigma_{h h }^{(\gamma=1)}(N)    
\end{equation}
\section{Signature of criticality in quantum spin chains}
Using the exact solution from \eqref{XYChainWithTransverseField} we can compute the 1-body reduced density matrix which requires computing the transverse magnetization $m_z(\gamma, h, N)$ at any real value of $h$ and $\gamma$. The results for XY and TFIM chains are obtained by putting $h=0$ and $\gamma=1$ in this full model, giving us $m_z(\gamma, N)$ and $m_z(h,N)$, respectively. The advantage of using the 1-body reduced density matrix here is the analytical simplicity of the susceptibility of entanglement entropy as:
\begin{equation}
    \Sigma_{N}(\lambda) = \frac{1}{2} \frac{(\partial_{\lambda}m_z(\lambda,N))^2}{1- m^2_z(\lambda,N)}, \ \ \ \ \text{when} \ \ \lambda \in \{ h, \gamma \}    \label{SusceptibilityOfentanglementEntropy}
\end{equation}
We can numerically determine its turning points as a function of $N$ which turns out to be at the global maxima of this susceptibility for any odd $N$ for the XY case and even $N$ for the TFIM case.

\begin{figure}[h!]
    \centering
    \begin{minipage}{0.49\textwidth}
        \centering
        \includegraphics[width=\textwidth, height=6.5cm, keepaspectratio]{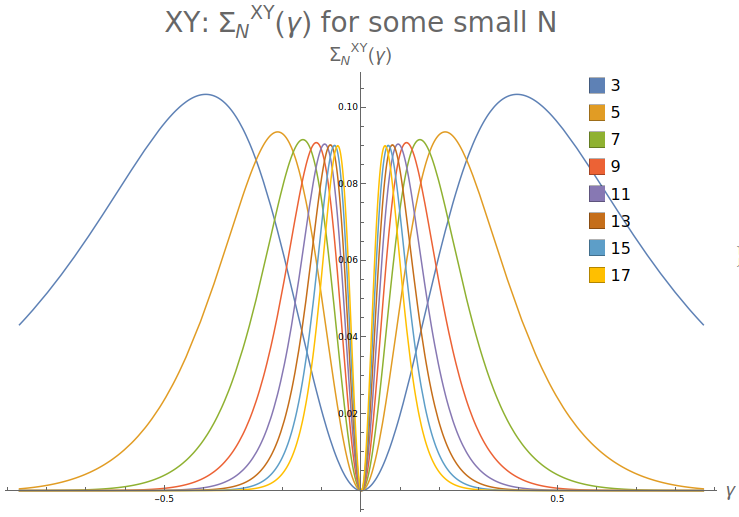}
        \label{fig:xy_suscept_many_small_size}
    \end{minipage}
    \hfill
    \begin{minipage}{0.49\textwidth}
        \centering
        \includegraphics[width=\textwidth, height=6.5cm, keepaspectratio]{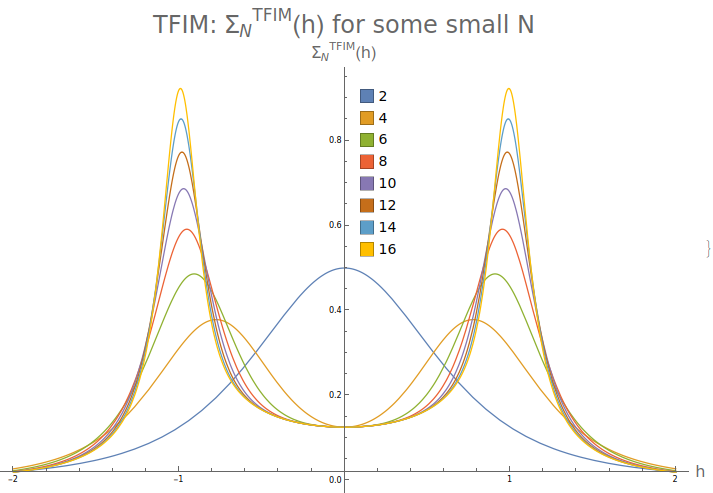}
    \end{minipage}
    \caption{Susceptibility of entanglement entropy 1. $ \Sigma_{N}^{\text{XY}}(\gamma) $,  2. $ \Sigma_{N}^{\text{TFIM}}(h) $ in a few small 1. XY and 2. TFIM chains}
    \label{fig:SusceptDataFor2DifferentModelsForSmallN}
\end{figure}

 \subsection{Finite XY chain}
\subsubsection{Numerical results}

There is no phase transition for any finite system as the susceptibility remains finite. For infinite system it identically diverges at the true critical point $\gamma=0$. The convergence of the turning points and the global maxima of $\Sigma_{N}^{\text{XY}}(\gamma)$ for any odd $N$ is demonstrated with the following figure (Fig \ref{fig:xy_plotsfordifferentsystemsizes}) where the symmetric turning points around $\gamma=0$ can be seen to converge to $\gamma_c^{\infty}=0$. while the maximum susceptibility can also be seen to saturate at a value of $\text{Max}(\Sigma^{\text{XY}}(\gamma)) \sim 0.0897653$.
 \begin{figure}[ht]
  \centering
  \includegraphics[width=\textwidth]{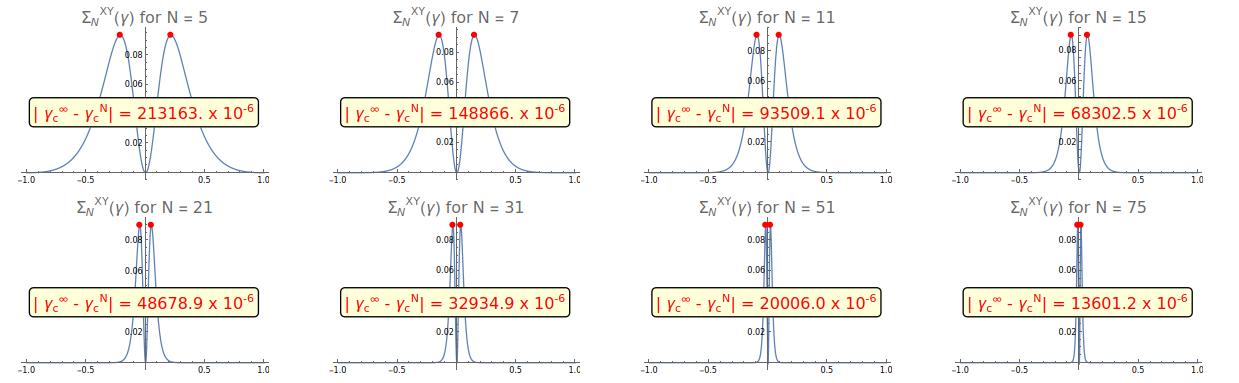}
  \caption{ Susceptibility of Entanglement-Entropy $\Sigma_{\gamma \gamma}^{(h=0)}$ for XY model}
  \label{fig:xy_plotsfordifferentsystemsizes}
\end{figure}

\begin{figure*}[ht]
    \centering
    \begin{minipage}{0.32\textwidth}
        \centering
        \includegraphics[width=\textwidth]{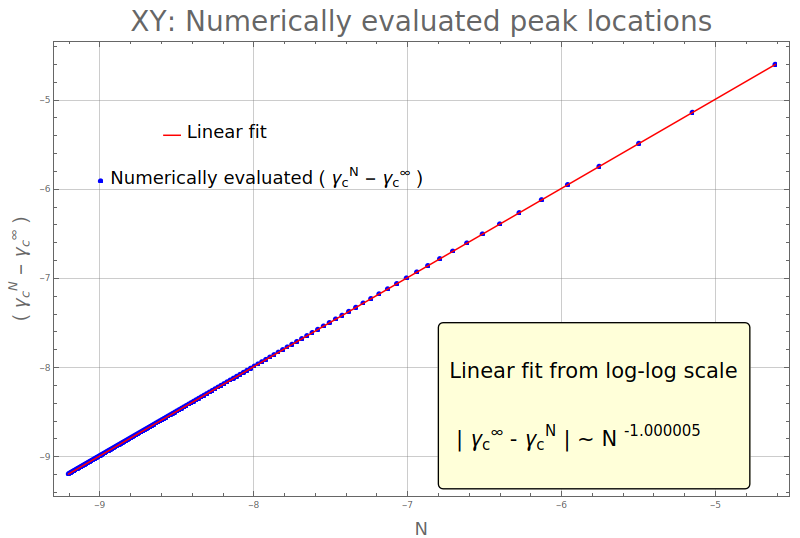}
    \end{minipage}
    \hfill
    \begin{minipage}{0.32\textwidth}
        \centering
        \includegraphics[width=\textwidth]{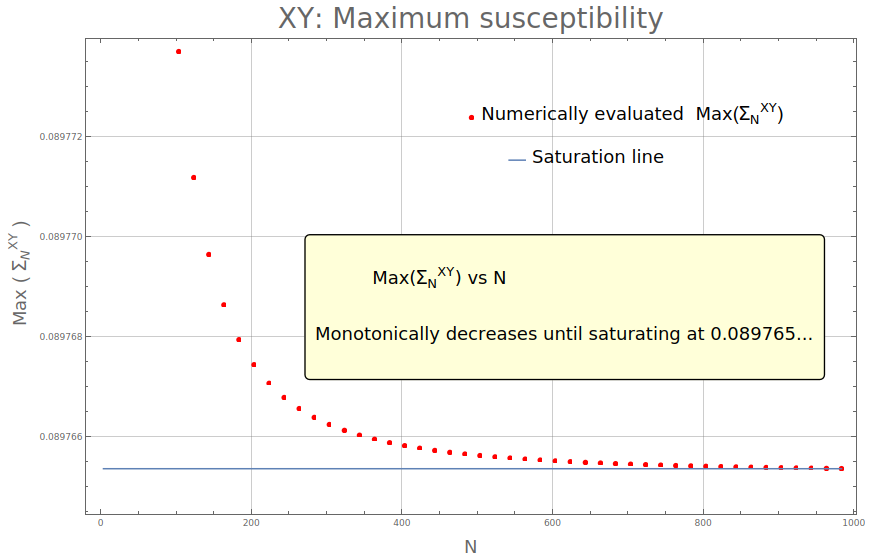}
    \end{minipage}
    \hfill
    \begin{minipage}{0.32\textwidth}
        \centering
        \includegraphics[width=\textwidth]{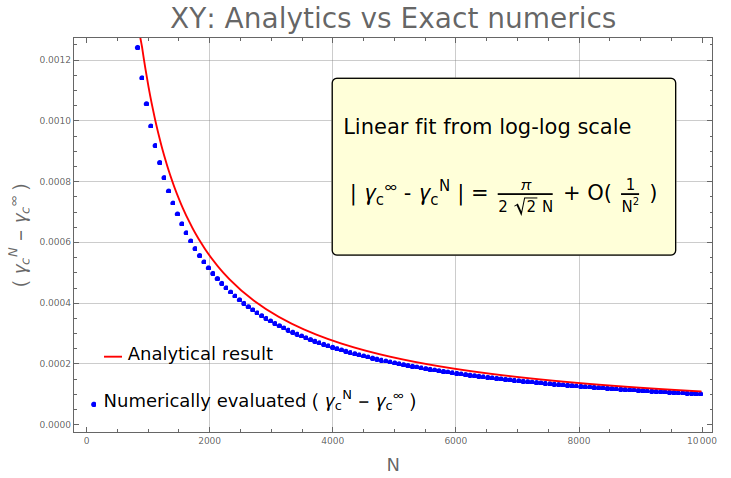}
    \end{minipage}
    \caption{1.  $| \gamma^{\infty}_c - \gamma^{ N}_c|$ in log-log scale, 2. Maximum susceptibility,
    3.Numerics vs analytics }
    \label{fig:ScalingDataPlotsInXY}
\end{figure*}

Not only the convergence of the peak to the true critical point, the narrowing of the susceptibility near the maxima directly indicates strongest and most concentrated susceptibility near $\gamma=0$ as the system size increases. In the log-log scale the distance between the turning points of this susceptibility from the true critical point (Fig \ref{fig:ScalingDataPlotsInXY}.1) is found to be a power-law in system size with an exponent $(\sim -1.000005)$. Its numerically evaluated values collapse in a straight line in the log-log scale without any noticable exception. The integer power law can be derived using perturbation theory as in the following way.

As evident from (Fig ~\ref{fig:SusceptDataFor2DifferentModelsForSmallN}.1 \& Fig \ref{fig:xy_plotsfordifferentsystemsizes}) the maxima of this susceptibility saturates to a finite value  after monotonically decreasing as system size increases (Fig \ref{fig:ScalingDataPlotsInXY}.2). This may be attributed to the lack of long-range order in isotropic $(\gamma=0)$ XY chain in periodic boundary \cite{Barouch1970StatMechI, Barouch1971StatMechII}. 

\subsubsection{Analytical result}
Considering large $N$ one can use a perturbation in $\mathcal{O}(\frac{1}{N})$ to compute $m_z(\gamma,N)$ \eqref{m_z__N} which can then be used to determine the susceptibility $\Sigma_{N}^{\text{XY}}(\gamma)$ \eqref{SusceptibilityOfentanglementEntropy}. This can be analytically solved for $\gamma$ to determine $\gamma_c(N)$. The details are in the appendix A. The fit of the exact numerics with the perturbative result is demonstrated in Fig \ref{fig:ScalingDataPlotsInXY}.3.


\begin{equation}
    |\gamma^{\infty}_c - \gamma^{N}_c| = \frac{\pi}{2 \sqrt{2} N} + \mathcal{O}(\frac{1}{N^2}) \ \ \ \text{as} \ N \to \infty
\end{equation}

This explains the integer power law with exponent $(-1)$ obtained from exact numerical results. Notice the poor fit in (Fig \ref{fig:ScalingDataPlotsInXY}.3.) for small systems as the perturbative analysis depends on taylor expansion around $\frac{1}{N} = 0$. This explains the better fit for larger system sizes.

\subsection{Finite TFIM}
\subsubsection{Numerical result}
\begin{figure}[ht]
  \centering
  \includegraphics[width=\textwidth]{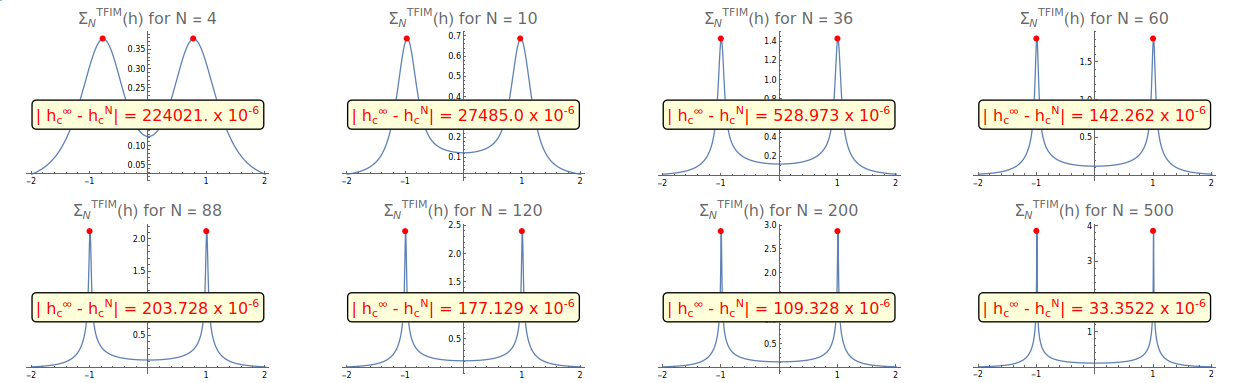}
  \caption{ Susceptibility of entanglement entropy $\Sigma_{N}^{\text{TFIM}}(h)$ for transverse field Ising model}
  \label{fig:TFIMSusceptForManySizes}
\end{figure}

There is no phase transition for any finite system as the susceptibility remains finite. For infinite system it identically diverges at the true critical points $h=\pm 1$. The similar behaviour for $\Sigma_{N}^{\text{TFIM}}(h)$ again rapidly converges to the true critical point in the following way, while narrowing sharply near the critical points (Fig \ref{fig:TFIMSusceptForManySizes}).

\begin{figure}[h!]
    \centering
    \begin{minipage}{0.32\textwidth}
        \centering
        \includegraphics[width=\textwidth]{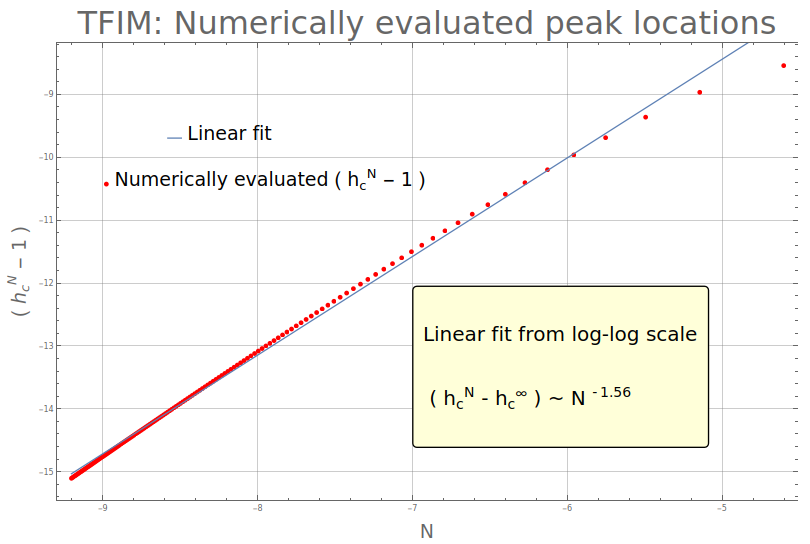}
    \end{minipage}
    \hfill
    \begin{minipage}{0.32\textwidth}
        \centering
        \includegraphics[width=\textwidth]{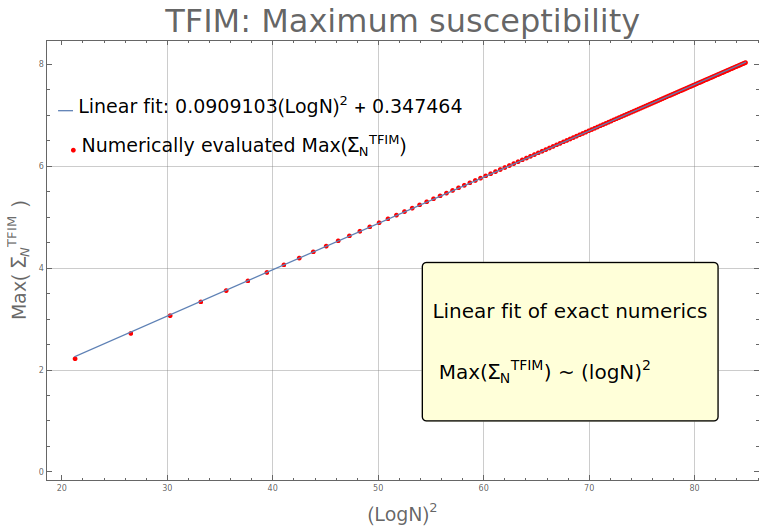}
    \end{minipage}
    \hfill
    \begin{minipage}{0.32\textwidth}
        \centering
        \includegraphics[width=\textwidth]{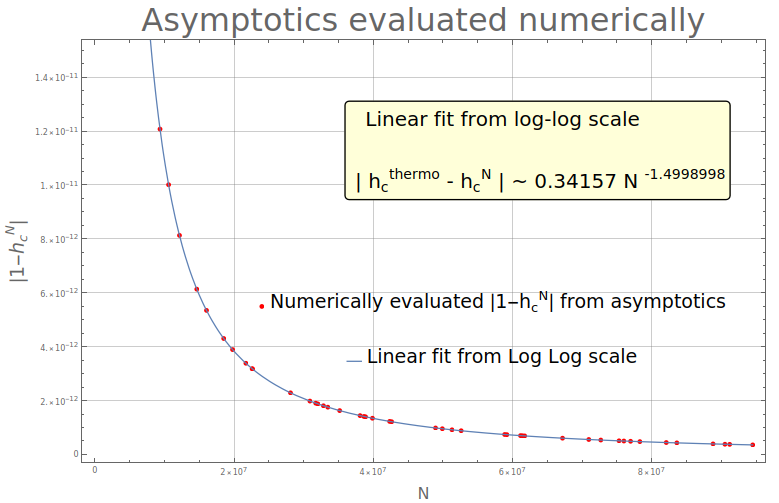}
    \end{minipage}

    \caption{1.  $(h^{N}_c - 1)$ in log-log scale , 2. Max($\Sigma_{N}^{\text{TFIM}}(h))$ vs $(\log N)^2$ , 3. Fit from asymptotics}
        \label{fig:TFIMAllScalingDataPlots}
        
\end{figure}

The 2 turning points near $h=\pm 1$ crosses $h= \pm 1$ at $N=48 \to 50$ from the side towards $h=0$ to the other, and eventually converges to $h = \pm 1$ from that side. This shows a power law with exponent $ \sim (-1.56)$. It is obtained by calculating slope from the linearly collapsed data $(h_c^N-h_c^\infty)$ in log-log scale (Fig \ref{fig:TFIMAllScalingDataPlots}.1).

As evident from (Fig ~\ref{fig:SusceptDataFor2DifferentModelsForSmallN}.2 \& Fig \ref{fig:TFIMSusceptForManySizes}) the maximum susceptibility scales as $\log(N)$ with an exponent $2$ (Fig \ref{fig:TFIMAllScalingDataPlots}.2). This may be attributed to the presence of long-range order in TFIM at the (pseudo-)critical points \cite{Barouch1970StatMechI, Barouch1971StatMechII, Sachdev2011QPT}.
\subsubsection{Analytical result}
For TFIM the ground state energy density $\epsilon(h, N)$ is related $m_z(h, N)$ as $\langle \hat{\sigma} \rangle = \partial_h \ \epsilon(h, N)$ with $\epsilon(h, N) = \frac{1}{N} \sum_{i=-\frac{N-1}{2}}^{i=\frac{N-1}{2}}  ( $ \\
$\sqrt{(h-\cos k)^2 + \sin^2 k} \ )$.
Which can be analysed through Jacobi theta function of 3rd kind, Chebychev's polynomials and generalised hypergeometric functions 3F2. It allows an asymptotic expression of the ground state energy density $\epsilon(h, N)$ when $N \to \infty$, as detailed in Appendix B. Giving us:
\begin{equation}
    \epsilon^{\text{asympt}}(h,N)=\frac{2(1+h)}{\pi} \ \text{E}(\frac{4h}{(1+h)^2}) - \ \frac{\sqrt{ \pi e} \ (h+1) }{ \ 2\pi \ N^{\frac{3}{2}} } \cdot (\frac{4h}{(1+h)^2})^N \cdot \Big( \ 1+\sqrt{1-\frac{4h}{(1+h)^2}} \ \Big)^{-2N}
\end{equation}
which we can use to compute $m_{z}^{\text{asympt}}(h,  N)$ and thus $\Sigma_{h}^{\text{asympt}}(N)$ as \eqref{asymptoticTFIM}. Due to the lack of general solution involving elliptic integrals, the final scaling is verified numerically from the asymptotic expression (in Fig \ref{fig:TFIMAllScalingDataPlots}.3), obtaining:
\begin{equation}
   | \  h_c^{\infty} - h_c^{N} \ | \sim \frac{1}{N^{\frac{3}{2}}} \ \ \ \ \text{as} \ \ N \to \infty
\end{equation}

This explains the numerically obtained power-law, as further verified with $0.34157 * N^{-1.4998998}$ as in Fig \ref{fig:TFIMAllScalingDataPlots}.3.

\section{Discussions}
This is directly generalisable for any n-site reduced density matrix \cite{Sachdev2011QPT, Barouch1970StatMechI, Barouch1971StatMechII} depending on the nature of the order parameter of the model under question, because all that is required is an one-to-one assignment of reduced density matrices at each point in the parameter space, which may be obtained through various numerical methods \cite{Tsai2001DMRG} and approximation techniques \cite{polkovnikov}. Notice that the single-site reduced density matrix used to analyse two cases of quantum criticality requires only the $\langle \sigma^z_i \rangle$ for any site $i$. Therefore given any arbitrarily large periodic TFIM (XY) chain with even (odd) number of spins $N$ one can estimate the size of the system by looking at the analytically obtained scaling fits. This means that with the information of the critical properties of this indicator one can experimentally observe any subsystem and still determine the range of entanglement across the whole system.  

Consider an arbitrary n-site reduced density matrix. Since entanglement is a measure of irreducibility of quantum dynamics of a system into any of its subsystem that is probabilistic in nature, so it has its own entropy. Now the proposed susceptibility measures how fast the relative entanglement entropy changes between that n-site and the rest of the system using a non-euclidean distance in the parameter space of the theory. This not only reveals a deeper geometric structure of quantum dynamics \cite{Kolodrubetz_2013, polkovnikov, Zanardi2006Fidelity, Zanardi2007Geometry}, but also gives a unique and model-agnostic way to calculate trajectories in parameter space that reduces the entanglement entropy globally, which may be useful in quantum control and large scale quantum technologies as optimal protocols \cite{polkovnikov}.

Further analysis is required to mathematically verify the divergence of maximum susceptibility but saturation of it, respectively for TFIM and XY chains and if at all the presence or absence of long-range order at the (pseudo-)critical points are related to it. Some calculations for quench and linear drive in TFIM suggest a decaying oscillatory behaviour of this susceptibility which is under investigation, indicating the utility of this tool even for dynamical quantum criticality \cite{Heyl2013DQPT, Sharma2016SlowQuenches, Sen2008Quenching, Zvyagin2016Review}.
\section{Conclusion}

A method of deriving the susceptibility of entanglement entropy of any given density matrix, has been proposed following the idea of Fisher-Rao metric in classical information geometry. Using the exact solution of one dimensional XY model in transverse field \cite{polkovnikov}, we have demonstrated the finite-size behaviours of turning points and global maximum of this susceptibility for an XY chain and a TFIM. The obtained power-law scaling of turning points of the aforementioned model has been verified using perturbative techniques, and a concoction of special functions respectively. 

This demands greater investigation of this approach, both for the pursuit of exact mathematical elegance and for the deeper understanding of quantum critical phenomenon, which has countless implications in modern quantum technologies.

\section{Acknowledgement}

I am truly grateful for the guidance of Prof. Arnab Sen at IACS, Kolkata as he not only directed me to the path I have appreciated most, leading to the idea of this work, but also encouraged me to pursue my scientific speculations and constantly helped to shape those concretely.

\printbibliography

\section{Appendices}
\subsection*{A. Turning point scaling in XY chain}

When $h=0$, the following summand becomes $g(\phi_l) = \frac{-cos(\phi_l)}{\sqrt{cos(\phi_l)^2 + \gamma^2sin^2(\phi_l)}}$, which has the property that $g(\pi-\phi_l) = - g(\phi_l)$ so that for even spins, momentum modes for either $[0, \pi]$ and $[-\pi,0]$ pairwise cancel, giving $m_z(0, \gamma,2N)= 0, \ \forall 
 \ \text{integer} \ N$. That is why everything remains well defined and non-trivial as long as we keep odd system sizes for XY.
$$m_z(0, \gamma, N)=-\frac{1}{N} \sum^{l=\frac{N-1}{2}}_{l=-\frac{N-1}{2}}\frac{\cos(\frac{2\pi}{N}l)}{\sqrt{\cos^2(\frac{2\pi}{N}l)+\gamma^2\sin^2(\frac{2\pi}{N}l)}}=-\frac{1}{2N} \sum^{l=\frac{N-1}{2}}_{l=-\frac{N-1}{2}}\Psi(\frac{2\pi}{N}l), \ \ \text{with} \ \ \Psi(k)=\frac{\cos(k)}{\sqrt{\cos^2(k)+\gamma^2\sin^2(k)}}$$
$$=-\frac{1}{2N}\Psi(0)-\frac{1}{2N} \sum^{j=\frac{N-1}{2}}_{j=1}\Psi(k_j)=-\frac{1}{N}(1+K) $$
$$\text{with} \ \ K=\sum^{j=\frac{N-1}{2}}_{j=1} \Psi(k_j)=\sum^{j=\frac{N-1}{2}}_{j=1}\frac{2\cos(k_j)}{\sqrt{\cos^2(k_j)+\gamma^2\sin^2(k_j)}}. \ \ \ \text{Now considering} \ \  k_j = \frac{2\pi}{N}j$$
$$ K=\begin{cases} \text{case 1} : \ \ \sum^{(\frac{N-1}{4}-1)}_{i=0} \Big[ \Psi(k_{i}^{+})+\Psi(k_{i}^{-}) \Big] \ \ \text{with} \ \  k_i^{\pm}=\frac{2\pi}{N}(\frac{N-1}{4}\pm i) & \text{if } \ N \mod 4 =1 \\ \text{case 2} : \ \ \Psi(k_{\frac{N+1}{4}})+ \sum^{(\frac{N-3}{4})}_{j=1} \Big[ \Psi(k_{j}^{+})+\Psi(k_{j}^{-})  \Big] \ \ \text{with} \ \ k_{j}^{\pm}=\frac{2\pi}{N}(\frac{N+1}{4}\pm i) & \text{if } \ N \mod 4 =3 \end{cases}$$

Since $N$ is always an odd number here, there is only $2$ possible values of $N \mod 4$, which are $1$ and $3$. Also notice that with the prefactor $\frac{1}{N}$ and a sum over N terms, the $\mathcal{O}(\frac{1}{N^s})$ term of the combination of $\Psi(\frac{2\pi j}{N})$s would yield corresponding $\mathcal{O}(\frac{1}{N^s})$ contribution in $m_z(0, \gamma, N)$ for all positive integer $s$ . Here $i$ and $j$ are just used to distinguish the indices for the 2 cases. To see how this arrangement simplifies this setup, notice that $k_i^{+}+k_i^{-}=\pi-\frac{\pi}{N}$ for the case 1, and  $k_j^{+}+k_j^{-}=\pi+\frac{\pi}{N}$ for the case 2. Also keep in mind that $\Psi(k)=-\Psi(\pi-k)$. This will help us make the perturbative expansion in such a way that things simplify very quickly.

\textbf{CASE 1 : }
$$\Psi(k_{i}^{+})+\Psi(k_{i}^{-})  = \Psi(k_{i}^{+})+\Psi(\pi-\frac{\pi}{N}-k_{i}^{+}) =\Psi(k_{i}^{+})-\Psi(k_{i}^{+}+\frac{\pi}{N})$$

$$=\Psi(k_{i}^{+})-[\Psi(k_{i}^{+})+\Psi'(k_{i}^{+})\frac{\pi}{N}+\mathcal{O}(\frac{1}{N^2})] \approx -\Psi'(k_{i}^{+})\frac{\pi}{N}$$

$$\text{now:} \ \ \Psi'(k_{i}^{+})=\Psi'(\frac{2\pi}{N}(\frac{N-1}{4}+ i))=\Psi'(\frac{\pi}{2}+\frac{\pi}{N}(2i-\frac{1}{2})) \approx \Psi'(\frac{\pi}{2})+ \Psi''(\frac{\pi}{2})\frac{\pi}{N}(2i-\frac{1}{2})+\mathcal{O}(\frac{1}{N^2})$$$$\text{To make sure of $\mathcal{O}(\frac{1}{N})$ terms in K we need: } \ \ \Psi(k_{i}^{+})+\Psi(k_{i}^{-}) \approx -\Psi'(\frac{\pi}{2})\frac{\pi}{N}=\frac{2\pi}{\gamma N}, \ \ \forall \ i$$$$\implies K^{( 1)} \approx \sum^{(\frac{N-1}{4}-1)}_{i=0} \frac{2\pi}{\gamma N}=\frac{2\pi}{\gamma N}(\frac{N-1}{4}), \ \ \ \text{at} \ \mathcal{O}(\frac{1}{N})$$
$$m_z^{(1)} \approx \frac{\pi - N \pi - 2 N \gamma}{2 N^2 \gamma} \implies \Sigma_{\gamma}^{(1)} \approx\frac{(-1 + N) \gamma^2}{\gamma^2 \left(-\left(-1 + N\right) \pi^2\right) - 4 N \pi \gamma + 4 N^2 (1 + N) \gamma^2}$$
$$\implies  \ \ \text{for large } \ N, \text{the turning points are }\ \gamma_{c,1}^{N} = \{ \ \gamma_{c,1}^{N, +}, \gamma_{c,1}^{N,-} \ \} $$

\begin{equation}
    \approx \{ \ \frac{3 \pi}{8 N (1+N)} + \frac{1}{8} \sqrt{\frac{\pi^2 + 8 N^2 \pi^2}{N^2 (1+N)^2}} \ , \ \ \  \frac{3 \pi}{8 N (1+N)} - \frac{1}{8} \sqrt{\frac{\pi^2 + 8 N^2 \pi^2}{N^2 (1+N)^2}} \ \}
\end{equation}

Which gives us the 2 turning points around $\gamma_c^{\text{thermo}}=0$, so that $$\gamma^{N, \pm}_{c,1} = \pm\frac{\pi}{2\sqrt{2}N} +\frac{ \frac{3\pi}{8} \mp \frac{\pi}{2\sqrt{2}}}{N^2}+ \mathcal{O}\left(\frac{1}{N^3}\right)$$
$$\implies \ \ \text{At the leading order for large $N$ } \  : \ \ \  |\gamma^{N, \pm}_{c,1}-\gamma^{\text{thermo}}_c| \approx \frac{\pi}{2\sqrt{2}N}$$

\textbf{CASE 2 :} This case corresponds to $N$ such that $N \mod 4 =3$. 
$$\Psi(k_{\frac{N+1}{4}})=\Psi(\frac{\pi}{2}+\frac{\pi}{2N})=\cancelto{0}{\Psi(\frac{\pi}{2})}+\Psi'(\frac{\pi}{2})\frac{\pi}{2N}+\mathcal{O}(\frac{1}{N^2})\approx -\frac{ \pi}{\gamma N}$$

$$\text{and} \ \ \Psi(k_{j}^{+})+\Psi(k_{j}^{-})  =\Psi(k_{j}^{+})+\Psi(\pi+\frac{\pi}{N}-k_{j}^{+})= \Psi(k_{j}^{+})-\Psi(k_{j}^{+}-\frac{\pi}{N}) $$

$$=\Psi(k_{j}^{+})-[\Psi(k_{j}^{+})-\Psi'(k_{j}^{+})\frac{\pi}{N}+\mathcal{O}(\frac{1}{N^2})] \approx \Psi'(k_{j}^{+})\frac{\pi}{N}$$

$$\text{furthermore: } \ \Psi'(k_{j}^{+})=\Psi'(\frac{\pi}{2}+\frac{\pi}{N}(2j+\frac{1}{2}))=\Psi'(\frac{\pi}{2})+\Psi''(\frac{\pi}{2})\frac{\pi}{N}(2j+\frac{1}{2})+\mathcal{O}(\frac{1}{N^2})$$$$\text{Similarly to make sure of $\mathcal{O}(\frac{1}{N})$ terms in K we need: } \ \ \Psi(k_{j}^{+})+\Psi(k_{j}^{-}) \approx \Psi'(\frac{\pi}{2})\frac{\pi}{N}=-\frac{2\pi}{\gamma N}, \ \ \forall \ j$$$$\implies K^{(2)} \approx  -\frac{ \pi}{\gamma N}-\sum^{(\frac{N-3}{4})}_{j=1}\frac{2\pi}{\gamma N}=-\frac{ \pi}{\gamma N}-\frac{2\pi}{\gamma N}(\frac{N-3}{4})        , \ \ \ \text{at} \ \mathcal{O}(\frac{1}{N})$$

$$m_z^{(2)} \approx \frac{\pi - N\pi + 2N\gamma}{2N^2\gamma} \implies \Sigma_{\gamma}^{(2)} \approx \frac{(-1 + N) \pi^2}{\gamma^2 \left(-((-1 + N) \pi^2) + 4N\pi\gamma + 4N^2(1 + N)\gamma^2\right)}$$
$$\implies  \ \ \text{for large } \ N, \text{the turning points are }\ \gamma_{c, 2}^{N} = \{ \ \gamma_{c,2}^{N, +}, \gamma_{c,2}^{N,-} \ \} $$

\begin{equation}
    \approx \{ \ -\frac{3\pi}{8N(1+N)} + \frac{1}{8}\sqrt{\frac{\pi^2 + 8N^2\pi^2}{N^2(1+N)^2}}  \ , \ \ -\frac{3\pi}{8N(1+N)} - \frac{1}{8}\sqrt{\frac{\pi^2 + 8N^2\pi^2}{N^2(1+N)^2}}   \ \}
\end{equation}

Which gives us the 2 turning points around $\gamma_c^{\text{thermo}}=0$, now $$\gamma^{N, \pm}_{c,2} \approx \pm \frac{\pi}{2\sqrt{2}N} - \frac{(3 \pm 2\sqrt{2})\pi}{8N^2} + \mathcal{O}\left(\frac{1}{N^3}\right)$$
$$\implies \ \ \text{At the leading order for large $N$ } \  : \ \ \  |\gamma^{N, \pm}_{c,2}-\gamma^{\text{thermo}}_c| \approx \frac{\pi}{2\sqrt{2}N}$$

Which agrees for both the cases, so we can conclude : 
\begin{equation}
    |\gamma^{N}_{c}-\gamma^{\text{thermo}}_c| \approx \frac{\pi}{2\sqrt{2}N} \ \ \text{as} \ N \to \infty
\end{equation}

\subsection*{B. Ground state energy density of TFIM}
When $\gamma=1$, taking odd system size in Eq \eqref{m_z__N} inevitably gives a $k=0$, which corresponds to $(h-1)$ in the denominator, hitting one of the the thermodynamic critical points, something that we will avoid by restricting to only even system sizes for TFIM. Consider the ground state energy density of a finite TFIM with periodic boundary $ \epsilon(h, N)$.
$$\text{It is more general to start with : } \ \epsilon(h, N) =  \frac{1}{N} \sum_{l=-\frac{N-1}{2}}^{l=\frac{N-1}{2}}\sqrt{(h-\cos k_l)^2+\sin^2 k_l}$$
$$\partial_h \ \epsilon(h, N) = \frac{1}{N} \sum_{l=-\frac{N-1}{2}}^{l=\frac{N-1}{2}}\frac{h-\cos k_l}{\sqrt{(h-\cos k_l)^2+\sin^2 k_l}} = m_z(h,N)$$
$$\text{so focus on : } \ \  \epsilon(h, N) =   \frac{1}{N} \sum_{l=-\frac{N-1}{2}}^{l=\frac{N-1}{2}} \sqrt{(h-\cos (\frac{k_l}{N}))^2+\sin^2 (\frac{k_l}{N})}, \ \ \text{with} \ \ \ \ l_i = 2\pi l$$
$$\text{notice: } \ \sqrt{(h-\cos (\frac{k_l}{N}))^2+\sin^2 (\frac{k_l}{N})} = \int^{\pi}_{-\pi} \sqrt{(h-\cos x)^2+\sin^2x} \cdot \delta(x-\frac{k_l}{N}) \cdot dx$$
$$\implies \epsilon(h, N) =   \int^{\pi}_{-\pi} \sqrt{(h-\cos x)^2+\sin^2x} \cdot  \ \Big( \  \frac{1}{N} \sum_{l= -\frac{N-1}{2}}^{\frac{N-1}{2}}\delta(x-\frac{k_l}{N}) \ \Big) \  \cdot dx  $$
$$=  \int^{\pi}_{-\pi} \sqrt{(h-\cos x)^2+\sin^2x} \cdot f(x, N) \cdot dx \ \text{and : } \ \ f(x,  N) = \frac{1}{N} \sum_{l= -\frac{N-1}{2}}^{\frac{N-1}{2}}\delta(x-\frac{k_l}{N})$$$\text{Notice: } \   \delta(x-\frac{k_l}{N}) =   N \ \delta(Nx-k_l) \ \text{so } \ f(x,  N)  \ \text{can be expanded in Fourier basis } \  \{ e^{i k Nx}\}$ which means that we have $f(x,  N) = \sum_{k=-\infty}^{\infty}c_k \ e^{ \  ikNx}$ with the fourier coeffients 
$$c_k = \frac{1}{4 \pi} \int_{x}^{} e^{ \ - ikNx}   f(x, l_i, N) \ dx=           \frac{1}{4 \pi} \int_{x}^{}e^{ \ - ikNx}\Big( \  \frac{1}{N} \sum_{j= -\frac{N-1}{2}}^{\frac{N-1}{2}}\delta(x-\frac{k_j}{N}) \ \Big) \ dx = \frac{1}{4 \pi N}\sum_{j= -\frac{N-1}{2}}^{\frac{N-1}{2}} e^{  \ - ik \ k_j}$$$$ = \frac{1}{ 4\pi N}\sum_{j= -\frac{N-1}{2}}^{\frac{N-1}{2}} e^{  \ - ik \ (2\pi j ) }= \frac{1}{ 4\pi N}\sum_{j= -\frac{N-1}{2}}^{\frac{N-1}{2}} 1= \frac{N}{4\pi \ N} =\frac{1}{4 \pi}=c_{k}, \ \ \forall \ k. $$$$ \text{so} \ \ \  f(x, l_i, N) = \frac{1}{4\pi} \sum_{q=-\infty}^{\infty} \  e^{ \ iqNx} = \frac{1}{4\pi} \sum_{q=-\infty}^{\infty} (1)^{q^2}  e^{ \ 2 qi  (\frac{Nx}{2})} = \frac{1}{4\pi}\theta_{3}(\frac{Nx}{2}, 1). \  $$
$$\text{where } \ \theta_3(z,q) = \sum_{n=-\infty}^\infty q^{n^2}  e^{ \ 2niz} = 1+2\sum_{n=1}^{\infty}q^{n^2} \cos(2nz)$$
$$ \ \  \ \text{is the Jacobi theta function of 3-rd kind.} \implies \theta_3(\frac{Nx}{2},1) = 1+2\sum_{q=1}^{\infty} \cos(qNx)$$
$$\text{Now consider: } \ \sqrt{(h-\cos x)^2+\sin^2x} = \sum_{n=0}^{\infty}c_n(h) \  T_n(\cos{x}) $$
represented as a linear combination of Chebyshev polynomials with their orthonormality as :
$$\ \int^{1}_{-1}\frac{T_n(x)T_m(x)}{\sqrt{1-x^2}}dx = \int^{\pi}_{0}T_n(\cos{y})T_m(\cos{y}) \ dy = \int^{\pi}_{0}\cos{(n \ y)}\cos{(m \ y)} \ dy =\begin{cases}  \frac{\pi}{2}\delta_{mn}, & \forall \  m \ \& \  n\neq 0 \\ \pi, & \forall \ m \ or \ n = 0\end{cases} $$
$$\ \implies \ c_n(h) = \frac{2}{\pi} \int_{0}^{\pi}T_n(\cos{x})\sqrt{(h-\cos x)^2+\sin^2x}  \ dx=  \frac{2}{\pi} \int_{0}^{\pi}\cos{(n \ x)}\sqrt{1-2h\cos{x}+h^2}  \ dx $$
$$= 2(1+h) \cdot {}_{3}F^{(R)}_{2}[\{ -\frac{1}{2},\frac{1}{2}, 1 \}; \{ 1-n,1+n \}; \frac{4h}{(1+h)^2}] \ $$
Where ${}_{3}F^{(R)}_{2}$ is the regularised generalised hypergeometric function.
$$\implies \epsilon(h, N)  =  \int^{\pi}_{-\pi} \sqrt{(h-\cos x)^2+\sin^2x} \cdot f(x, l_i, N) \cdot dx =  2 \int^{\pi}_{0} \Big( \ \sum_{n=0}^{\infty}c_n(h) \  T_n(\cos{x}) \ \Big) \cdot \frac{1}{4\pi} \ \theta_{3}(\frac{Nx}{2}, 1) \cdot dx$$
$$ =     \frac{1}{2\pi} \cdot \sum_{n=0}^{\infty}  c_n(h) \ \Big[  \ \int^{\pi}_{0}  \cos{(n \ x)} \ \theta_{3}(\frac{Nx}{2}, 1)  \cdot dx \ \Big] =   \frac{1}{2\pi} \cdot \sum_{n=0}^{\infty}  c_n(h)  \cdot V_{n}(N) , \ \ \ \ with \ \ \  \theta_3(Nx,1) = 1+2\sum_{q=1}^{\infty} \cos(qNx),$$
$$  \text{we get} \ \ \ V_{n}(N) = \int^{\pi}_{0}  \cos{(n \ x)} \ \theta_{3}(\frac{Nx}{2}, 1) \ dx $$
$$= \int^{\pi}_{0}  \cos{(n \ x)} \Big [ \ 1+2\sum_{q=1}^{\infty} \cos(qNx) \ \Big] dx =  \int^{\pi}_{0}  \cos{(n \ x)} \ dx \ + \ 2\sum_{q=1}^{\infty}\int^{\pi}_{0}  \cos{(n \ x)}\cos(qN \ x) \ dx $$
$$\text{when \ } \ n=0, \ V_0(N) = \pi, \ \ \ \& \ \forall n>0, \in \mathbb{N} \ \ we \ get$$
$$ V_{n>0}(N)=  \frac{\cancel{\sin{(n  \pi)}}}{n} + \ 2\sum_{q=1}^{\infty} \ G(n, qN), \ \ \ \ with \ \ a, b \in \mathbb{N}, \ \text{we havea} \ $$
$$G(a, b) = \int^{\pi}_{0}  \cos{(a  x)}\cos(bx) \ dx  = \begin{cases}  \frac{\pi}{2}\delta_{ab}, & \forall \  a \ \& \  b\neq 0 \\ \pi, & \forall \ a \ or \ b = 0 \end{cases} \ \ \ \ \ \text{from orthogonality of } \ T_{n}(\cos{x})'s$$
$$ So \ when \ \ \{ q, N, n \} >0, \  \&\in \mathbb{N}, \ \ \ G( q N, n)  = \frac{\pi}{2}\delta_{n, \ qN} \  \ \implies  \ V_0(N) = \pi \ \ \ \& \ \ \ \ V_{n>0}(N) =  \pi \sum_{q=1}^{\infty} \delta_{n, \ qN}   $$
   $$\implies \epsilon(h, N)  =  \frac{1}{2\pi} \cdot  \sum_{n=0}^{\infty}  c_n(h) \cdot V_n{(N)} = \frac{1}{2\pi} \Big[ \ c_0(h)V_0(N) + \sum_{n=1}^{\infty}  c_n(h)  \cdot   V_{n>0}{(N)} \ \Big] $$
   $$= \frac{1}{2\pi} \Big[ \ c_0(h) \pi + \sum_{n=1}^{\infty}  c_n(h)  \cdot   ( \ \pi\sum_{q=1}^{\infty} \delta_{n, \ qN}   \ ) \ \Big] =\frac{1}{2} \Big[c_0(h) + \sum_{q=1}^{\infty}c_{qN}(h) \Big]  \equiv \frac{1}{2}\sum_{q=0}^{\infty}c_{qN}(h)$$
  \begin{equation}
        \ \implies \epsilon(h, N)  = \frac{1}{2} \cdot \sum_{q=0}^{\infty} c_{qN}(h)  = (1+h) \cdot \sum_{q=0}^{\infty}{}_{3}F^{(R)}_{2}[\{ -\frac{1}{2},\frac{1}{2}, 1 \}; \{ 1-qN,1+qN \}; \frac{4h}{(1+h)^2}]  \ 
   \end{equation}
This is the exact ground state energy density of finite 1D TFIM.   
$$\text{for our context: } \ m_z(h, N)= \partial_h \ \epsilon(h, N), \ \Sigma(h, N) = \frac{\partial^2_h\ \epsilon(h, N)}{1-(\partial_h \ \epsilon(h, N))^2}, \ \ \ \ \ \ \text{so ( with } \ n \in \mathbb{N} \  \text{) focus on : } \ \ $$
 $${}_{3}F^{(R)}_{2}[\{ -\frac{1}{2},\frac{1}{2}, 1 \}; \{ 1-n,1+n \};  \ z] = \frac{1}{\Gamma(1-n)\Gamma(1+n)} \ {}_{3}F_{2}[\{ -\frac{1}{2},\frac{1}{2}, 1 \}; \{ 1-n,1+n \}; \ z]$$
Now we will use the property of our hypergeometric function from \href{https://functions.wolfram.com/HypergeometricFunctions/Hypergeometric3F2/03/07/02/0001/}{\text{wolfram documentation link}}.
$$ \begin{aligned} {}_3F_2\left[ \  \{ \begin{array}{c} a, b, c \}; \ \{ a - n, e \end{array} \}; \  z \ \right] = \frac{1}{(1-a)_n} \sum_{k=0}^n \frac{(-1)^k (1-a)_{n-k} (b)_k (c)_k}{(e)_k} \  \binom{n}{k} \ {}_2F_1[ b+k, c+k; e+k; z]  \ z^k \end{aligned} $$
$$\text{with} \ b=\frac{1}{2}, c=-\frac{1}{2}, e = 1+n,\ \& \ a \to 1 \ \text{for this case}$$
 $$\text{note} \ \  \lim_{x\to 0} (x)_{n-k} = \frac{\Gamma(n-k)}{\Gamma(0)} = \delta_{nk}, \ \ \text{because} \ \Gamma(0)=\infty$$
  So the only contribution comes from $k=n$.
 $$\lim_{a\to 1} {}_3F^{(R)}_2\left[ \  \{ \begin{array}{c} a, b, c \}; \ \{ a - n, e \end{array} \}; \  z \ \right] $$
$$=\lim_{a\to 1} \Big[ \ \frac{1}{\Gamma(a-n)\Gamma(e)(1-a)_n} \sum_{k=0}^n \frac{(-1)^k (1-a)_{n-k} (b)_k (c)_k}{(e)_k} \  \binom{n}{k} \ {}_2F_1[ b+k, c+k; e+k; z]  \ z^k   \ \Big] $$
 $$\lim_{a\to 1}\frac{1}{\Gamma(a-n)\Gamma(1+n)(1-a)_n} = \lim_{a\to 1}\frac{\Gamma(1-a)}{\Gamma(a-n)\Gamma(1+n) \Gamma(n+1-a)} =\lim_{a\to 1}\frac{ \Gamma(1-a)\sin(\pi(a-n))}{\Gamma(1+n)  \ \pi\ } $$
 $$\text{Now}, \ \ \sin(\pi(a-n))= \sin(\pi a)\cos(\pi n)-\sin(\pi n )\cos(\pi a)= \sin(\pi a) = - \sin(\pi - \pi a) =  - \sin( \ \pi(1 -  a) \ ), $$
 $$\text{because $n$ is even, } \ \implies  \lim_{a\to 1} \Gamma(1-a)\sin(\pi(a-n)) = \pi$$
$$\implies \lim_{a\to 1}\frac{ \Gamma(1-a)\sin(\pi(a-n))}{\Gamma(1+n)  \ \pi\ } = \frac{1}{\Gamma(1+n)}$$
  $$\implies \lim_{a\to 1} {}_3F^{(R)}_2\left[ \  \{ \begin{array}{c} a, \frac{1}{2}, - \frac{1}{2} \}; \ \{ a - n, 1+n \end{array} \}; \  z \ \right] =\lim_{a\to 1} \Big[ \ \frac{1}{\Gamma(a-n)\Gamma(1+n)(1-a)_n} \sum_{k=0}^n $$
$$\frac{(-1)^k (1-a)_{n-k} (\frac{1}{2})_k (-\frac{1}{2})_k}{(1+n)_k} \  \binom{n}{k} \ {}_2F_1[ k+\frac{1}{2}, \ k-\frac{1}{2}; \ 1+n+k; \ z]  \ z^k   \ \Big] $$
 $$= \frac{1}{\Gamma(1+n)} \sum_{k=0}^n \frac{(-1)^k \  (\frac{1}{2})_k (-\frac{1}{2})_k \ \delta_{nk}  }{(1+n)_k} \  \binom{n}{k} \ {}_2F_1[ k+\frac{1}{2}, \ k-\frac{1}{2}; \ 1+n+k; \ z]  \ z^k$$
$$ =\frac{1}{\Gamma(1+n)} \frac{ (\frac{1}{2})_n (-\frac{1}{2})_n   }{(1+n)_k} \   {}_2F_1[ n+\frac{1}{2}, \ n-\frac{1}{2}; \ 2n+1; \ z]  \ z^n, \ \ \ \text{as n is even}  $$

 $$\ {}_3F^{(R)}_2\left[ \  \{ \begin{array}{c} 1, \frac{1}{2}, - \frac{1}{2} \}; \ \{ 1 - n, 1+n \end{array} \}; \  z \ \right]=z^n  \cdot\frac{  (\frac{1}{2})_n (-\frac{1}{2})_n   }{\Gamma(1+n) \ (1+n)_n} \cdot   {}_2F_1[ n+\frac{1}{2}, \ n-\frac{1}{2}; \ 2n+1; \ z]  $$

$$\text{also note that}\ \ \ \frac{  (\frac{1}{2})_n (-\frac{1}{2})_n   }{\Gamma(1+n) \ (1+n)_n} = \frac{\Gamma(n+\frac{1}{2})\Gamma(n-\frac{1}{2} )\Gamma(1+n) \ }{\Gamma(1+n) \Gamma(2n+1) \ \Gamma(\frac{1}{2}) \ \Gamma(-\frac{1}{2})}, \ \ \ \text{and} \ \ \frac{1}{\Gamma(\frac{1}{2})\Gamma(-\frac{1}{2})}=-\frac{1}{2 \pi} $$
 $$\implies {}_3F^{(R)}_2\left[   \{ \begin{array}{c} 1,  \frac{1}{2},   - \frac{1}{2} \};  \{ 1 - n,  1+n \end{array} \};   z  \right]=  -  z^n \cdot \frac{  \Gamma(n+\frac{1}{2})\Gamma(n-\frac{1}{2})}{2 \pi  \Gamma(2n+1)}    \cdot   {}_2F_1[ n+\frac{1}{2},  n-\frac{1}{2}; \ 2n+1;  z] $$
And we want
$$ c_{2qN}(h)=2 \ (1+h) \ {}_3F^{(R)}_2\left[  \{ \begin{array}{c} 1, \frac{1}{2},  - \frac{1}{2} \};  \{ 1 - 2qN,  1+2qN \end{array} \};  \frac{4h}{(1+h)^2}  \right]$$
 $$\implies \epsilon(h, N)  = \frac{1}{2} \cdot \sum_{q=0}^{\infty} c_{qN}(h) = \epsilon_{\text{thermo}}(h) + \epsilon_{\text{finite}}(N, h)$$
 
 $$\text{with : } \ c_{qN}(h)= - \ (1+h) (\frac{4h}{(1+h)^2})^{qN} \cdot \frac{ \Gamma(qN+\frac{1}{2})\Gamma(qN-\frac{1}{2})}{2 \pi \ \Gamma(2qN+1)}    \cdot   {}_2F_1[ qN+\frac{1}{2}, \ qN-\frac{1}{2}; \ 2qN+1; \frac{4h}{(1+h)^2}],$$
 $$ \epsilon_{\text{thermo}}(h)= \frac{c_0(h)}{2}, \ \ \ \  \epsilon_{\text{finite}}(N, h)=\frac{1}{2} \cdot \sum_{q=1}^{\infty} c_{qN}(h), \ \ \ \& \ \ \ \epsilon_{\text{finite}}^{(p)}(N, h):=\frac{1}{2} \cdot \sum_{q=1}^{p} c_{qN}(h).$$The reason to do this is that for large n, $(\frac{4h}{(1+h)^2})^n \cdot \frac{  \Gamma(n+\frac{1}{2})\Gamma(n-\frac{1}{2})}{2 \pi  \Gamma(2n+1)}    \cdot   {}_2F_1[ n+\frac{1}{2},  n-\frac{1}{2}; \ 2n+1;  \frac{4h}{(1+h)^2}]$ has the following properties,

\begin{enumerate}
    \item  this particular $|\frac{4h}{(1+h)^2}|\leq 1, \  \forall h \in \mathbb{R}$ so the prefactor is finite and $\leq 1$, and the 2F1 converges for all n.

    \item the prefactor containing gamma functions vanishes faster than exponential with n, i.e. as $ \frac{1}{e^{n} \ n^{\frac{s}{2}}} \sim e^{- \frac{s
    }{2} n \log{n}}$ for all positive integers $n, s$ as : $$\frac{  \Gamma(n+\frac{1}{2})\Gamma(n-\frac{1}{2})}{2 \pi  \Gamma(2n+1)} \sim e^{-2\ n \\ \log2} \left[ \frac{1}{2\sqrt{\pi}} \left(\frac{1}{n}\right)^{\frac{1}{2}} + \frac{3}{16\sqrt{\pi}} \left(\frac{1}{n}\right)^{\frac{5}{2}}  +  \ ... \right]\to 0  \ as \ n \to \infty.$$

    \item the ${}_2F_1[ n+\frac{1}{2},  n-\frac{1}{2}; \ 2n+1;  (\frac{4h}{(1+h)^2})]$ is finite everywhere for all h and n, except for large n it is 0 everywhere except kinks near $h \sim \pm 1$. 
\end{enumerate}

This immediately tells that for large N, leading contribution in $\epsilon(h, N)$ comes from $q=0$ which gives us $\epsilon_{\text{thermo}}( h)$ and is not a function of size because the limit has been taken already; so for next correction when N is sufficiently large but finite we can consider $q=1$ to get the leading N dependence, which is encapsulated inside $\epsilon_{\text{finite}}(N, h)$. This way one can consider $p$ many terms in the expansion for intermediate system sizes where $\epsilon_{\text{finite}}^{(p)}(N, h)$ may be helpful. Now,

\begin{equation}
    \epsilon_{\text{thermo}}( h) = -(1+h) \frac{ \ \Gamma(\frac{1}{2})\Gamma(-\frac{1}{2})}{2 \pi \ \Gamma(1)} \cdot {}_2F_1[\frac{1}{2},-\frac{1}{2}; \ 1; \ \frac{4h}{(1+h)^2}] = \frac{2(1+h)}{\pi} \ \text{E}(\frac{4h}{(1+h)^2}) \ \ \& 
    \end{equation}
\begin{equation}\implies m_{z, \ \text{thermo}}(h) = \partial_h \ \epsilon_{\text{thermo}}( h) = 
\frac{(h+1)}{h \pi \ } \ \text{E}(\frac{4h}{(1+h)^2}) + \frac{(h-1)}{h \pi \ } \ \text{K}(\frac{4h}{(1+h)^2})
\end{equation}

Here $E(x)$ is the complete elliptic integral of second kind, and $K(x)$ is the complete elliptic integral of first kind. These are exactly with the results when analytically integrated for thermodynamically large system. Now asymptotics of the next leading terms are needed as functions of N and h, which as we can see, would give fractional powers of N. I suppose this will also carry to the distance of our desired peaks from $\pm 1$ because there is a simple pattern for differentiating any pFq's.

If this is correct then if there is any intermediate system size where certain behaviors change, it must be contained within this series of hypergeometric functions as this is exact and analytical.

Now let us obtain the leading contribution in $\epsilon(h, N) = \epsilon_{thermo}(h) + \epsilon_{N}(h)$, i.e. by keeping only $q=0 \ \& \ 1$ terms for $N \to \infty$. Consider $h \geq 0$ and $g_h = \frac{4h}{(1+h)^2}$, which means $$\epsilon_{N}(h) = -(h+1) \ g_h^{N} \ \frac{\Gamma(N+\frac{1}{2})\Gamma(N-\frac{1}{2})}{2\pi \ \Gamma(2N+1)} \cdot {}_2F_{1}[ \ N-\frac{1}{2}, N+\frac{1}{2}; \ 2N+1; \ g_h \ ]$$
$$\text{notice here:} \ \ {}_2F_{1}[ \ N-\frac{1}{2}, N+\frac{1}{2}; \ 2N+1; \ z \ ] = \sum_{k=0}^{\infty} \frac{(N-\frac{1}{2})_k (N+\frac{1}{2})_k}{(2N+1)_k}\frac{z^k}{k!}, \ \ \text{where} \ \ $$
$$(N-\frac{1}{2})_k = \frac{\Gamma(N+k-\frac{1}{2})}{\Gamma(N-\frac{1}{2})}=\frac{\Gamma(N+k)}{\Gamma(N)} \cdot \frac{\Gamma(N)}{\Gamma(N-\frac{1}{2})} \cdot \ \frac{\Gamma(N+k-\frac{1}{2})}{\Gamma(N+k)}=  \frac{(N)_k \ \Gamma(N)}{\Gamma(N-\frac{1}{2})} \cdot \ \frac{\Gamma(N+k-\frac{1}{2})}{\Gamma(N+k)} $$
$$\text{furthermore:} \ \ \frac{\Gamma(N+k-\frac{1}{2})}{\Gamma(N+k)} \sim (N+k)^{-\frac{1}{2}}, \ \ \text{which as} \ \ N \to \infty \ \ \text{can be written as }$$

$$\frac{\Gamma(N+k-\frac{1}{2})}{\Gamma(N+k) \ k!} \sim \frac{1}{\sqrt{N}}\implies (N-\frac{1}{2})_k \sim \frac{(N)_k \ \Gamma(N)}{\Gamma(N-\frac{1}{2}) \ \sqrt{N}}$$

$$\implies {}_2F_{1}[ \ N-\frac{1}{2}, N+\frac{1}{2}; \ 2N+1; \ z \ ] = \sum_{k=0}^{\infty} \frac{(N-\frac{1}{2})_k (N+\frac{1}{2})_k}{(2N+1)_k}\frac{z^k}{k!}=$$
$$\frac{ \Gamma(N)}{\Gamma(N-\frac{1}{2}) \ \sqrt{N}} \cdot \sum_{k=0}^{\infty} \frac{(N)_k (N+\frac{1}{2})_k}{(2N+1)_k}\frac{z^k}{k!} \sim \frac{ \Gamma(N)}{\Gamma(N-\frac{1}{2}) \ \sqrt{N}}\cdot{}_{2}F_{1}[ \ N, N+\frac{1}{2}; \ 2N+1; \ z]$$

$$\implies {}_2F_{1}[ \ N-\frac{1}{2}, N+\frac{1}{2}; \ 2N+1; \ z \ ] \sim \frac{ \Gamma(N)}{\Gamma(N-\frac{1}{2}) \ \sqrt{N}} \cdot \Big( \ \frac{1}{2}+\frac{1}{2}\sqrt{1-z} \ \Big)^{-2N}, \ \ \ \ \text{as} \ \ N \to \infty.$$ from identity 15.4.17 in \href{https://dlmf.nist.gov/15.4}{NIST}.
Also note that $$ \frac{\Gamma(N+\frac{1}{2})\Gamma(N-\frac{1}{2})}{\Gamma(2N+1)} \to \frac{\sqrt{2 \pi e}}{N \cdot 2^{2N+\frac{1}{2}}}, \ \text{as} \ N\to \infty$$
Finally we obtain for large $N$, 
$$\epsilon_{N}(h) \sim -(h+1) \ g_h^{N} \cdot  \frac{\sqrt{2 \pi e}}{2\pi \ 2^{2N+\frac{1}{2}} N^{\frac{3}{2}} } \cdot \Big( \ \frac{1}{2}+\frac{1}{2}\sqrt{1-g_h} \ \Big)^{-2N} $$

$$= - \ \frac{\sqrt{ \pi e} \ (h+1) }{ \ 2\pi \ N^{\frac{3}{2}} } \cdot (\frac{4h}{(1+h)^2})^N \cdot \Big( \ 1+\sqrt{1-\frac{4h}{(1+h)^2}} \ \Big)^{-2N}$$

This will be used as leading contribution of finite size effects in $\epsilon(h, N)$ when $N$ is large. Therefore asymptotically we have 

\begin{equation}
    \epsilon^{\text{asympt}}(h,N)=\frac{2(1+h)}{\pi} \ \text{E}(\frac{4h}{(1+h)^2}) - \ \frac{\sqrt{ \pi e} \ (h+1) }{ \ 2\pi \ N^{\frac{3}{2}} } \cdot (\frac{4h}{(1+h)^2})^N \cdot \Big( \ 1+\sqrt{1-\frac{4h}{(1+h)^2}} \ \Big)^{-2N}
\end{equation}

Using this the asymptotic expressions of $m_z(h, N)$ and correspondingly $\Sigma_{h}(N)$ is derived as  

\begin{equation}
    m_{z}^{\text{asympt}}(h,  N)=\partial_{h}\epsilon^{\text{asympt}}(h,N), \ \ \ \& \ \ \ \Sigma_{h}^{\text{asympt}}(N)= \frac{1}{2}\frac{(\partial_{h}m_{z}^{\text{asympt}}(h,  N))^2}{1-(m_{z}^{\text{asympt}}(h,  N))^2}
    \label{asymptoticTFIM}
\end{equation}

Recall that we are only interested in the turning points of $\Sigma_{h}^{\text{asympt}}(N)$ and how it changes as $N \to \infty$. It is clear how long the expression is going to be but since solving the zeros of its derivative essentially means solving an equation with of the kind $f(x)+g_1(x)E(h(x))+g_2(x)K(h(x))=0$, which does not have any known general solution because $E$ and $K$ here are respectively complete elliptic integrals of second and first kind. So we have presented the numerical convergence of its peak towards $h_c^{\infty}=1$. 

Due to the presence of the elliptic integral $K(x)$ the function is numerically unstable near $x=1$ although regularity is found when $N \in [10^5, 10^8]$, working precision 80 and accuracy goal 48 have been considered to evaluate the turning points of this $\Sigma_{h}^{\text{asympt}}(N)$.

\subsection*{C. Mathematica notebook:}

To reproduce every diagrammatic results presented in this work the reader is requested to visit \href{https://drive.google.com/drive/folders/1OOkHZZ-GNMqXay8YgeABE3bYWuJNc8Sa?usp=sharing}{this link}. It contains the detailed mathematica notebook to compute every presented data.

\end{document}